\documentclass[a4paper,fleqn]{cas-dc}

\usepackage[numbers,square,sort&compress]{natbib}
\def\tsc#1{\csdef{#1}{\textsc{\lowercase{#1}}\xspace}}
\tsc{WGM}
\tsc{QE}

\begin{document}
\let\WriteBookmarks\relax
\def\floatpagepagefraction{1}
\def\textpagefraction{.001}

\shorttitle{}

\shortauthors{J. Jia et al.}

\title[mode=title]{Remote nonreciprocal magnon–magnon entanglement via the Barnett effect with coherent feedback}

\author[1,2,3,4]{Jinhao Jia}[orcid=0009-0001-5756-032X]
\author[1,2,3]{Yingru Li}[orcid=0009-0007-8218-4649]
\author[1,2,3]{Ran Liang}
\author[1,2,3]{Qinghua Cai}
\author[4]{Xin Wen}

\author[1,2,3]{Mei Zhang}[orcid=0000-0002-9728-9001]
\cormark[1]


\affiliation[1]{
    organization={School of Physics and Astronomy, Beijing Normal University},
    city={Beijing},
    postcode={100875},
    country={China}
}

\affiliation[2]{
    organization={Applied Optics Beijing Area Major Laboratory, Beijing Normal University},
    city={Beijing},
    postcode={100875},
    country={China}
}

\affiliation[3]{
    organization={Key Laboratory of Multiscale Spin Physics (Ministry of Education), Beijing Normal University},
    city={Beijing},
    postcode={100875},
    country={China}
}

\affiliation[4]{
	organization={Beijing Academy of Quantum Information Sciences},
	city={Beijing},
	postcode={100193},
	country={China}
}

\cortext[1]{Corresponding author}

\begin{abstract}
	We propose a scheme for generating and controlling remote
	nonreciprocal magnon--magnon entanglement in a cavity magnonic system with coherent feedback.The two microwave cavities are driven by independent
single-mode squeezed vacuum fields, with a rotating
YIG sphere in the left cavity and a stationary one
in the right cavity.
For a fixed rotation direction, reversing the bias
magnetic field applied to the rotating sphere changes
the sign of the Barnett shift, leading to nonreciprocal
magnon--magnon entanglement.
By tuning the cavity detuning and feedback phase,
entanglement can be generated for one direction of
the bias magnetic field while it vanishes for the
opposite direction, yielding complete nonreciprocity.
	Coherent feedback enhances the entanglement at an
	appropriate reflectivity, whereas stronger feedback
	can reduce it.
	The relative squeezing phase provides additional control,
	and an appropriate choice of unequal squeezing factors
	improves the entanglement at fixed total input photon
	occupation.
	With the squeezed inputs kept fixed, moderate feedback
	also increases the critical temperature for magnon
	entanglement.
	Our results provide a controllable approach to preparing
	remote nonreciprocal quantum correlations in cavity
	magnonic systems.
\end{abstract}


\begin{highlights}
\item The Barnett effect enables remote nonreciprocal magnon-magnon entanglement.
\item Cavity detuning and feedback phase enable complete nonreciprocity.
\item Feedback can enhance entanglement without increasing its nonreciprocity.
\item Unequal squeezing can enhance entanglement at fixed total input photon occupation.
\item Moderate feedback improves thermal robustness with fixed squeezed inputs.
\end{highlights}



\maketitle 
\section{Introduction}\label{sec0}

Quantum entanglement~\cite{Horodecki2009,Laure2021,Shlomi2021,Tao2020,lv2018} is a fundamental feature of quantum mechanics and an essential resource for quantum communication~\cite{Ursin2007,Piveteau2022} and quantum metrology~\cite{Pezz2018,zhang2015}, and quantum computation~\cite{Graham2022,Bartolucci2023}.
Cavity magnonics provides a promising platform for
generating and controlling macroscopic quantum
entanglement~\cite{Tabuchi2014,Hisatomi2016,Dany2017,Wolski2020,Babak2022}.
Magnons in yttrium iron garnet (YIG) spheres represent
collective spin excitations, whose high spin density and
low damping facilitate strong coupling to microwave
photons \cite{Huebl2013,Tabuchi2014}.
The magnon frequency can also be tuned by an external
bias magnetic field, providing a convenient means of
controlling the system dynamics.
These properties make cavity magnonics a promising
platform for the generation of macroscopic quantum
entanglement via magnetostrictive interactions~\cite{Li2018,Yuan2025,Li2019,Zhang2016},
magnon Kerr nonlinearity~\cite{Zhang2019,Guo2025,Ji2023,Xiong2022,Zhang2021}, and squeezed microwave fields~\cite{Yu2020,Zhang2022,Lu2024,peng2025}.

Besides generating entanglement, controlling its
nonreciprocal behavior provides additional possibilities
for manipulating quantum correlations~\cite{Hu2026,Imara2016,Yang2026,Chen2023,Chen2024}.
The Barnett effect describes the magnetization induced
by mechanical rotation and establishes a connection
between rotational motion and spin angular momentum~\cite{Ono2015,Kani2022,Bretzel2009}.
In a rotating YIG sphere, this effect produces a shift
of the magnon frequency.
For a fixed rotation direction, reversing the bias
magnetic field changes the sign of the Barnett shift,
allowing quantum correlations to depend on the magnetic
field direction.
Nonreciprocal entanglement and quantum coherence have
been investigated in cavity magnomechanical systems
via the Barnett effect~\cite{Lu2025,Jia2026}.
These studies show that quantum correlations can be
controlled by reversing the direction of the bias
magnetic field.

Coherent feedback provides another approach to controlling
quantum correlations \cite{Lloyd2000,Schmid2022}.
In a coherent feedback loop, the output field is returned
to the system without measurement.
The amplitude reflectivity and feedback phase determine
the returned field and affect the effective cavity decay,
detuning, and coupling to input noise.
Coherent feedback has been explored for manipulating quantum states and enhancing quantum correlations, with applications including optical squeezing~\cite{Wiseman1994,Gough2009}, control of mechanical resonators~\cite{Sudhir2017,Kerckhoff2013,Harwood2021}, quantum error correction~\cite{Kerckhoff2010,Kerckhoff2011,Cramer2016}, and entanglement enhancement in both optical~\cite{Asjad2016,Li2017,Peng2023,Zhou2024,Jia2026Coherent,Jiao2026} and microwave fields~\cite{Amazioug2025,Liu2025,Gao2025,Lin2025}.

The above developments motivate investigating how
coherent feedback affects remote entanglement in the
presence of the Barnett effect.
Remote magnon--magnon entanglement has been proposed
using quantum-correlated microwave fields
\cite{Yu2020}.
More recently, Gong et al. showed that interference
between two independent squeezed vacuum fields enhances
entanglement between magnons in coupled microwave
cavities \cite{Gong2026Remote}.
For such a system, an important question is how coherent
feedback influences both the entanglement strength and
its nonreciprocity when a magnon frequency is modified
by the Barnett effect.
Since feedback changes both the cavity decay and its
coupling to the squeezed input, reducing the effective
decay rate alone does not determine the resulting
entanglement.
Moreover, enhancing entanglement and increasing its
nonreciprocity are distinct objectives.
It is therefore necessary to examine whether they can
be controlled simultaneously through the feedback
parameters and the squeezed inputs.

In this work, we propose a scheme for generating and
controlling remote nonreciprocal magnon--magnon
entanglement in a cavity magnonic system with
coherent feedback.
Two microwave cavities are connected by a coaxial cable,
with a rotating YIG sphere in the left cavity and a
stationary one in the right cavity.
Each cavity is driven by an independent squeezed vacuum
field, and a coherent feedback loop is connected to the
right cavity.
We compare the magnon entanglement
for opposite bias magnetic field directions while
keeping the other parameters fixed.
We show that feedback enhances the entanglement at an
appropriate reflectivity, although this enhancement
does not necessarily increase its normalized contrast.
By tuning the cavity detuning and feedback phase,
entanglement can exist for one magnetic field direction
and vanish for the opposite direction, yielding complete
nonreciprocity.
The relative squeezing phase and the two squeezing factors
provide further control.
Finally, with the squeezed inputs kept fixed, moderate
feedback increases the critical temperature for
magnon--magnon entanglement.
These results provide a controllable scheme for preparing
remote nonreciprocal quantum correlations in cavity
magnonic systems.

\section{System and model}\label{sec1}

\begin{figure*}[t]
	\centering
	\includegraphics[width=0.98\textwidth]
	{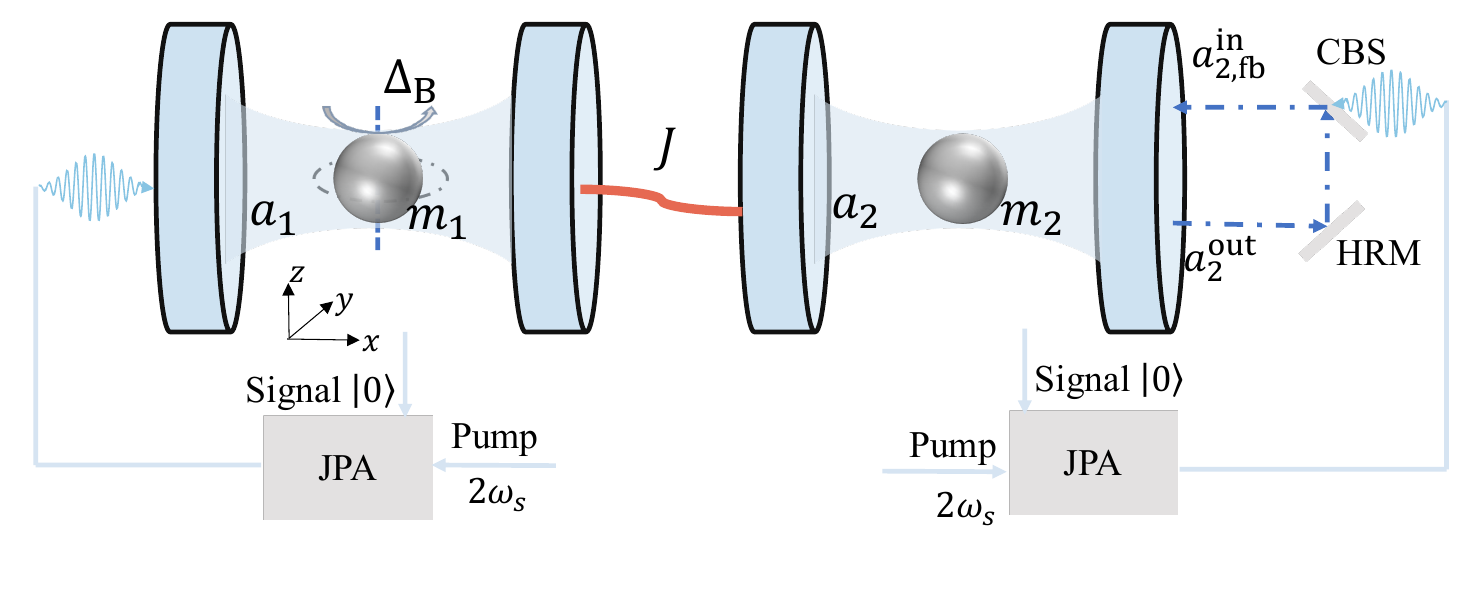}
	\caption{Schematic of the proposed system. Two microwave cavities, coupled via a coaxial cable with strength \(J\), support photon modes \(a_j\) and contain YIG spheres with magnon modes \(m_j\) (\(j=1,2\)). The left sphere rotates about the \(z\) axis, inducing a Barnett shift \(\Delta_B\), while the right sphere remains stationary. Two JPAs, pumped at \(2\omega_s\) with vacuum signal inputs, generate independent squeezed vacuum fields centered at \(\omega_s\). The right-cavity output \(a_2^{\mathrm{out}}\) is reflected by a HRM, mixed with the second squeezed input at a CBS, and reinjected as \(a_{2,\mathrm{fb}}^{\mathrm{in}}\).}
	\label{fig1}
\end{figure*}

As shown in Fig.~\ref{fig1}, the system consists of two spatially separated microwave cavities connected by a coaxial cable, with a rotating YIG sphere in the left cavity and a stationary one in the right cavity. Each cavity supports a microwave photon mode with frequency $\omega_{aj}$, while its embedded YIG sphere supports a magnon mode representing collective spin excitations~\cite{Gong2026Remote,Li2018}. The bare resonance frequency of the \(j\)th magnon mode is given by \(\omega_{mj}=\gamma H_j\), where \(H_j\) is the magnitude of the corresponding bias magnetic field and \(\gamma\) is the gyromagnetic ratio (\(j=1,2\))~\cite{Zhang2016}. Each magnon mode couples to the photon mode through the magnetic dipole interaction. For the YIG sphere rotating around the $z$ axis with angular frequency $\Delta_B$, the Barnett effect produces an emergent magnetic field $H_B=\Delta_B/\gamma$~\cite{Ono2015,Kani2022,Bretzel2009}. This field induces a magnetization proportional to $\Delta_B$, whose polarity is determined by the rotation direction. Because of angular momentum conservation, the frequency of the magnon mode shifts from $\omega_{m1}$ to $\omega_{m1}+\Delta_B$. Therefore, when the YIG sphere is kept rotating counterclockwise at a fixed angular frequency, the Barnett shift of the magnon mode can change sign by reversing the direction of the bias magnetic field along the $z$ axis.
The two microwave cavities are driven by independent single-mode squeezed vacuum fields generated by two Josephson parametric amplifiers (JPAs). Each JPA is pumped at frequency \(2\omega_s\), with its signal input kept in vacuum, producing a squeezed microwave vacuum field centered at \(\omega_s\)~\cite{Yu2020,Zhang2022,Lu2024,peng2025}. The squeezing parameters and phases of the two fields are denoted by \(R_j\) and \(\theta_j\), respectively. The first squeezed field is injected directly into the left cavity, while the second enters the right cavity through a controllable beam splitter (CBS).
The right cavity is connected to a coherent feedback loop formed by the CBS and a highly reflecting mirror (HRM)~\cite{Asjad2016,Li2017,Peng2023,Zhou2024,Jia2026Coherent,Jiao2026}. The cavity output field \(a_2^{\mathrm{out}}\) is reflected by the HRM and coherently mixed with the second squeezed vacuum field at the CBS before being fed back into the cavity as \(a_{2,\mathrm{fb}}^{\mathrm{in}}\)~\cite{Jia2026Coherent,Jiao2026}. The amplitude reflectivity and transmissivity of the CBS are denoted by \(r_b\) and \(t_b\), respectively, satisfying \(r_b^2+t_b^2=1\).  \(\theta\) denotes the total feedback-loop phase at frequency \(\omega_s\). We assume a lossless feedback loop and neglect its propagation delay.

In a frame rotating at \(\omega_s\), the system Hamiltonian under the rotating-wave approximation reads~\cite{Li2019}

\begin{equation}
  \begin{aligned}
	H/\hbar={}&
	\sum_{j=1}^{2}
	\left[
	\Delta_{aj}a_j^\dagger a_j
	+\Delta_{mj}m_j^\dagger m_j
	+g_j\left(a_j^\dagger m_j+a_jm_j^\dagger\right)
	\right]\\
	&+\Delta_{\mathrm B}m_1^\dagger m_1
	+J\left(a_1^\dagger a_2+a_1a_2^\dagger\right).
   \end{aligned}
\label{eq1}
\end{equation}
Here, \(\Delta_{aj}=\omega_{aj}-\omega_s\) and \(\Delta_{mj}=\omega_{mj}-\omega_s\). \(a_j\) (\(a_j^\dagger\)) and \(m_j\) (\(m_j^\dagger\)) denote the cavity and magnon annihilation (creation) operators, respectively, satisfying \([a_j,a_k^\dagger]=[m_j,m_k^\dagger]=\delta_{jk}\).

Using the standard input--output relation, the right-cavity output is
$a_2^{\mathrm{out}}=\sqrt{2\kappa_{a2}}\,a_2
-\widetilde a_{2,\mathrm{fb}}^{\mathrm{in}}$.
The feedback input is given by
$\widetilde a_{2,\mathrm{fb}}^{\mathrm{in}}
=r_b e^{i\theta}a_2^{\mathrm{out}}+t_b a_2^{\mathrm{in}}$.
Combining these relations, we obtain~\cite{Jia2026Coherent}
\begin{equation}
	\widetilde a_{2,\mathrm{fb}}^{\mathrm{in}}
	=\frac{\sqrt{2\kappa_{a2}}\,r_b e^{i\theta}}
	{1+r_b e^{i\theta}}\,a_2
	+\frac{t_b}{1+r_b e^{i\theta}}\,a_2^{\mathrm{in}}.
	\label{eq:cfmm_feedback_field}
\end{equation}
The effective input noise is
$a_{2,\mathrm{fb}}^{\mathrm{in}}
=t_b a_2^{\mathrm{in}}/(1+r_b e^{i\theta})$.
Substituting the self-consistent feedback field into the cavity equation,
we obtain the quantum Langevin equations (QLEs)
\begin{equation}
	\begin{aligned}
		\dot a_1={}&-(i\Delta_{a1}+\kappa_{a1})a_1
		-ig_1m_1-iJa_2
		+\sqrt{2\kappa_{a1}}\,a_1^{\mathrm{in}},\\
		\dot a_2={}&-(i\Delta_{\mathrm{fb}}+\kappa_{\mathrm{fb}})a_2
		-ig_2m_2-iJa_1
		+\sqrt{2\kappa_{a2}}\,a_{2,\mathrm{fb}}^{\mathrm{in}},\\
		\dot m_j={}&-(i\widetilde{\Delta}_{mj}+\kappa_{mj})m_j
		-ig_ja_j+\sqrt{2\kappa_{mj}}\,m_j^{\mathrm{in}},
	\end{aligned}
	\label{eq:cfmm_langevin}
\end{equation}
where
\begin{equation}
	\begin{aligned}
		\Delta_{\mathrm{fb}}
		&=\Delta_{a2}
		-\frac{2\kappa_{a2}r_b\sin\theta}
		{1+r_b^2+2r_b\cos\theta},\\
		\kappa_{\mathrm{fb}}
		&=\kappa_{a2}
		\frac{1-r_b^2}{1+r_b^2+2r_b\cos\theta}
	\end{aligned}
	\label{eq:cfmm_effective_parameters}
\end{equation}
are the effective right-cavity detuning and decay rate, respectively~\cite{Harwood2021}.
Here, $\widetilde{\Delta}_{m1}=\Delta_{m1}+\Delta_B$ and
$\widetilde{\Delta}_{m2}=\Delta_{m2}$.
$\kappa_{aj}$ and $\kappa_{mj}$ denote decay rates of the photon mode and the magnon mode, respectively. $m_j^{\mathrm{in}}$ represents  input noise of the magnon mode.

The input noise is characterized by the following correlation functions:
\begin{equation}
	\begin{aligned}
		\langle a_j^{\mathrm{in}\dagger}(t)a_k^{\mathrm{in}}(t')\rangle
		&=N_j\delta_{jk}\delta(t-t'),\\
		\langle a_j^{\mathrm{in}}(t)a_k^{\mathrm{in}\dagger}(t')\rangle
		&=(N_j+1)\delta_{jk}\delta(t-t'),\\
		\langle a_j^{\mathrm{in}}(t)a_k^{\mathrm{in}}(t')\rangle
		&=M_j\delta_{jk}\delta(t-t'),\\
		\langle a_j^{\mathrm{in}\dagger}(t)a_k^{\mathrm{in}\dagger}(t')\rangle
		&=M_j^*\delta_{jk}\delta(t-t'),\\
		\langle m_j^{\mathrm{in}\dagger}(t)m_k^{\mathrm{in}}(t')\rangle
		&=\bar n_j\delta_{jk}\delta(t-t'),\\
		\langle m_j^{\mathrm{in}}(t)m_k^{\mathrm{in}\dagger}(t')\rangle
		&=(\bar n_j+1)\delta_{jk}\delta(t-t').
	\end{aligned}
	\label{eq:cfmm_thermal_correlations}
\end{equation}
with $N_j=\sinh^2R_j$ and $M_j=e^{i\theta_j}\sinh R_j\cosh R_j$. 
$\bar n_j=[\exp(\hbar\omega_{mj}/k_BT)-1]^{-1}$ is the mean thermal excitation number, where $k_B$ is the Boltzmann constant and $T$ is the environment temperature.

We decompose each mode operator into its steady-state mean value
and a quantum fluctuation:
\begin{equation}
	a_j=\langle a_{j}\rangle+\delta a_j,\qquad
	m_j=\langle m_{j} \rangle+\delta m_j,
\end{equation}
Since the squeezed and thermal inputs have zero mean, the stable
steady-state mean fields satisfy $\langle a_{j}\rangle=\langle m_{j}\rangle=0$.
Substituting these decompositions into the QLEs yields exact linear
equations for the fluctuations.

We introduce the quadrature fluctuation operators $\delta X_o=(\delta o+\delta o^\dagger)/\sqrt{2},\delta Y_o=(\delta o-\delta o^\dagger)/(\sqrt{2} \mathrm{i}),$
for $o=a_j,m_j$. The equations for the quadrature fluctuations can
then be written in matrix form,
\begin{equation}
	\dot{\boldsymbol u}(t)=A\boldsymbol u(t)+\boldsymbol n(t),
	\label{eq:cfmm_quadrature_dynamics}
\end{equation}
where$\boldsymbol u^{\mathrm T}(t)
	=[\delta X_{a1},\delta Y_{a1},
	\delta X_{a2},\delta Y_{a2},
	\delta X_{m1},\delta Y_{m1},
	\delta X_{m2},\delta Y_{m2}],$
and the input-noise vector is
\[
\begin{aligned}
	\boldsymbol n(t)=\bigl[
	&\sqrt{2\kappa_{a1}}X_{a1}^{\mathrm{in}},
	\sqrt{2\kappa_{a1}}Y_{a1}^{\mathrm{in}},
	\sqrt{2\kappa_{a2}}X_{a2,\mathrm{fb}}^{\mathrm{in}},
	\sqrt{2\kappa_{a2}}Y_{a2,\mathrm{fb}}^{\mathrm{in}},\\
	&\sqrt{2\kappa_{m1}}X_{m1}^{\mathrm{in}},
	\sqrt{2\kappa_{m1}}Y_{m1}^{\mathrm{in}},
	\sqrt{2\kappa_{m2}}X_{m2}^{\mathrm{in}},
	\sqrt{2\kappa_{m2}}Y_{m2}^{\mathrm{in}}
	\bigr]^T.
\end{aligned}
\]
The drift matrix is given by
\begingroup
\setlength{\arraycolsep}{3pt}
\begin{equation}
	A=
	\begin{pmatrix}
		-\kappa_{a1} & \Delta_{a1}
		& 0 & J & 0 & g_1 & 0 & 0\\
		-\Delta_{a1} & -\kappa_{a1}
		& -J & 0 & -g_1 & 0 & 0 & 0\\
		0 & J
		& -\kappa_{\mathrm{fb}} & \Delta_{\mathrm{fb}}
		& 0 & 0 & 0 & g_2\\
		-J & 0
		& -\Delta_{\mathrm{fb}} & -\kappa_{\mathrm{fb}}
		& 0 & 0 & -g_2 & 0\\
		0 & g_1 & 0 & 0
		& -\kappa_{m1} & \widetilde{\Delta}_{m1} & 0 & 0\\
		-g_1 & 0 & 0 & 0
		& -\widetilde{\Delta}_{m1} & -\kappa_{m1} & 0 & 0\\
		0 & 0 & 0 & g_2 & 0 & 0
		& -\kappa_{m2} & \widetilde{\Delta}_{m2}\\
		0 & 0 & -g_2 & 0 & 0 & 0
		& -\widetilde{\Delta}_{m2} & -\kappa_{m2}
	\end{pmatrix},
	\label{eq:cfmm_drift_matrix}
\end{equation}
\endgroup

The covariance matrix is defined by
	$V_{pq}(t)=\frac{1}{2}\langle u_p(t)u_q(t)+u_q(t)u_p(t)\rangle,$
with $p,q=1,\ldots,8$. Its evolution obeys
\begin{equation}
	\dot V(t)=AV(t)+V(t)A^{\mathrm T}+D,
	\label{eq:cfmm_covariance_evolution}
\end{equation}
where the diffusion matrix is determined by$\frac{1}{2}\langle n_p(t)n_q(t')+n_q(t')n_p(t)\rangle=D_{pq}\delta(t-t')$.
It has the block-diagonal form $D=D_{a1}\oplus D_{a2}\oplus D_{m1}\oplus D_{m2}$,
where $D_{a1}=\kappa_{a1}\mathsf S(N_1,M_1)$,
		$D_{a2}=\kappa_{\mathrm{fb}}\mathsf S(N_2,M_{2,\mathrm{fb}}),
		D_{mj}=\kappa_{mj}(2\bar n_j+1)I_2,$
and
\begin{equation}
	\nonumber \mathsf S(N,M)=\begin{pmatrix}
		2N+1+2\operatorname{Re}M&2\operatorname{Im}M\\
		2\operatorname{Im}M&2N+1-2\operatorname{Re}M
	\end{pmatrix}.
\end{equation}
For positive decay rates and $0\le r_b<1$, all eigenvalues of $A$ have
negative real parts. The steady-state covariance matrix therefore follows
from the Lyapunov equation
\begin{equation}
	AV+VA^{\mathrm T}=-D.
	\label{eq:cfmm_lyapunov}
\end{equation}

To quantify entanglement between the two magnon modes, we extract the
$4\times4$ submatrix of $V$ associated with
$(X_{m1},Y_{m1},X_{m2},Y_{m2})$,
\begin{equation}
	V_{mm}=\begin{pmatrix}
		V_{m1}&C_{mm}\\
		C_{mm}^{\mathrm T}&V_{m2}
	\end{pmatrix},
	\label{eq:cfmm_magnon_covariance}
\end{equation}
where $V_{m1}$ and $V_{m2}$ describe the individual magnon modes and
$C_{mm}$ contains their correlations. The logarithmic negativity is
\cite{VidalWerner2002,Plenio2005}
\begin{equation}
	E_N=\max\{0,-\ln(2\widetilde\nu_-)\},
	\label{eq:cfmm_logarithmic_negativity}
\end{equation}
where $\widetilde\nu_-$ is the smallest symplectic eigenvalue of the
partially transposed magnon covariance matrix~\cite{Simon2000}. Explicitly,
\begin{equation}
	\widetilde\nu_-=
	\sqrt{\frac{\widetilde\Sigma-
			\sqrt{\widetilde\Sigma^2-4\det V_{mm}}}{2}},
	\label{eq:cfmm_partial_transpose_eigenvalue}
\end{equation}
with $\widetilde\Sigma=\det V_{m1}+\det V_{m2}-2\det C_{mm}$.
The
vacuum variance is $1/2$, and the two magnons are entangled when
$\widetilde\nu_-<1/2$.
\section{Numerical simulations}\label{sec2}
In this section, we investigate the steady-state entanglement
between the two remote magnon modes. We first examine the effects
of the Barnett shift and coherent feedback on $E_N$, and then
quantify its asymmetry under reversal of $\Delta_B$. We further
study the roles of cavity detuning, feedback phase, and squeezing
parameters, and assess the robustness of the entanglement
against thermal noise.

We take experimentally feasible parameters
within the range of typical values~\cite{Zhang2016,Li2018,Li2019}:
$\kappa_{a1}/2\pi=5~\mathrm{MHz}$,
$\omega_s/2\pi=10~\mathrm{GHz}$,
$\kappa_{a2}=4\kappa_{a1}$,
$\kappa_{m1}=\kappa_{m2}=0.2\kappa_{a1}$,
$g_1=g_2=2\kappa_{a1}$,
$J=4\kappa_{a1}$,
$\Delta_{a1}=\Delta_{a2}=-4\kappa_{a1}$,
$\Delta_{m1}=-\Delta_{m2}=2\kappa_{a1}$,
$R_1=R_2=0.9$,
$\theta_1=\theta_2=\theta=0$, and $T=20~\mathrm{mK}$.
The stability of each solution is verified by requiring all
eigenvalues of the drift matrix to have negative real parts.

\begin{figure}[t]
	\centering
	\includegraphics[width=0.5\textwidth]
	{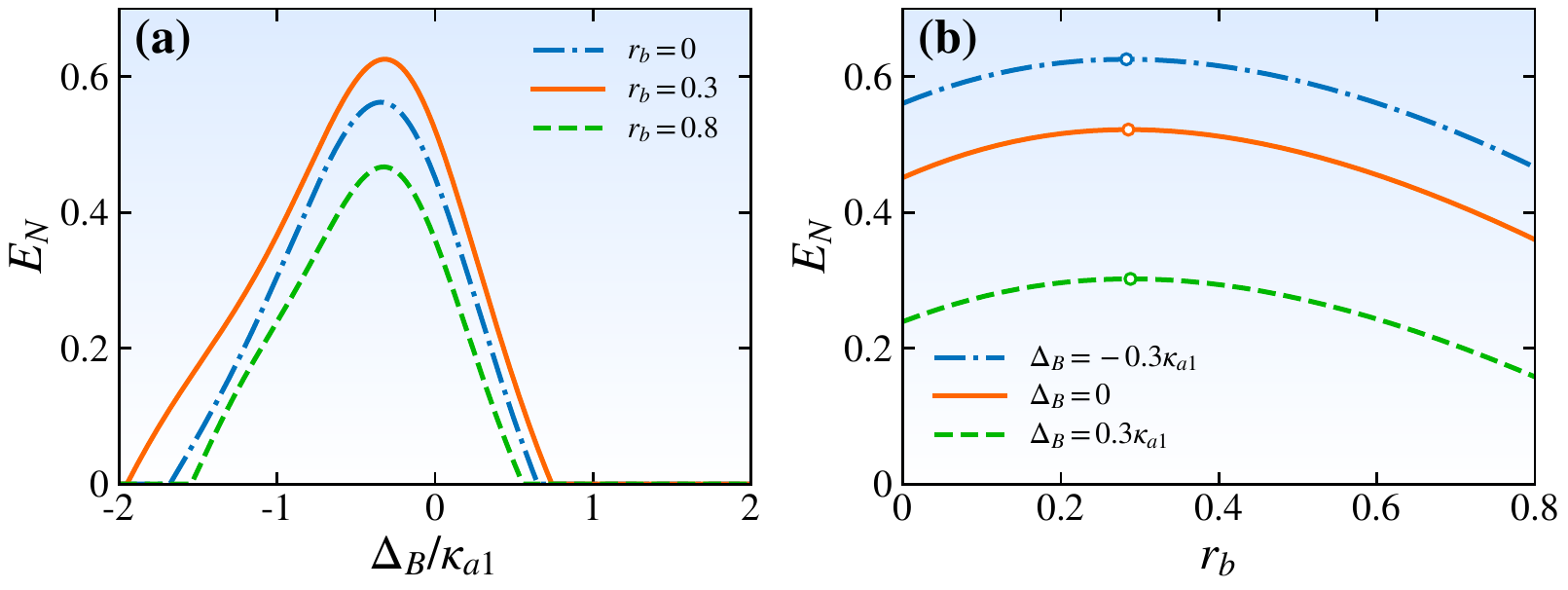}
	\caption{Steady-state magnon--magnon entanglement $E_N$ versus
		(a) the Barnett shift $\Delta_B/\kappa_{a1}$ for
		$r_b=0,0.3,0.8$, and (b) the feedback reflectivity $r_b$ for
		$\Delta_B/\kappa_{a1}=-0.3,0,0.3$.
		Open circles mark the maxima in panel (b).
		The other parameters are specified in the text.}
	\label{fig2}
\end{figure}

Figure~\ref{fig2}(a) shows the steady-state magnon--magnon
entanglement $E_N$ as a function of the Barnett shift $\Delta_B$
for different feedback reflectivities $r_b$.
The curves are asymmetric about $\Delta_B=0$, with the maximum
entanglement occurring at a negative Barnett shift.
The Barnett effect changes the detuning of $m_1$ and thus affects
the transfer of quantum correlations from the squeezed fields
to the magnon modes.
For the parameters considered here, opposite rotation directions
yield different values of $E_N$, demonstrating nonreciprocal
magnon--magnon entanglement.
Figure~\ref{fig2}(b) reveals a nonmonotonic dependence of $E_N$
on $r_b$.
For $\theta=0$, increasing $r_b$ simultaneously reduces the
effective cavity decay rate and the coupling of cavity $a_2$
to the squeezed input field.
The reduced decay rate lengthens the cavity lifetime and
initially enhances $E_N$, whereas the reduced coupling to
the squeezed field limits the transfer of quantum correlations
to the magnons.
At sufficiently large $r_b$, the latter effect becomes dominant,
and $E_N$ decreases.
The competition between these two effects therefore leads
to an optimal feedback reflectivity.

\begin{figure*}[htbp]
	\centering
	\includegraphics[width=0.98\textwidth]
	{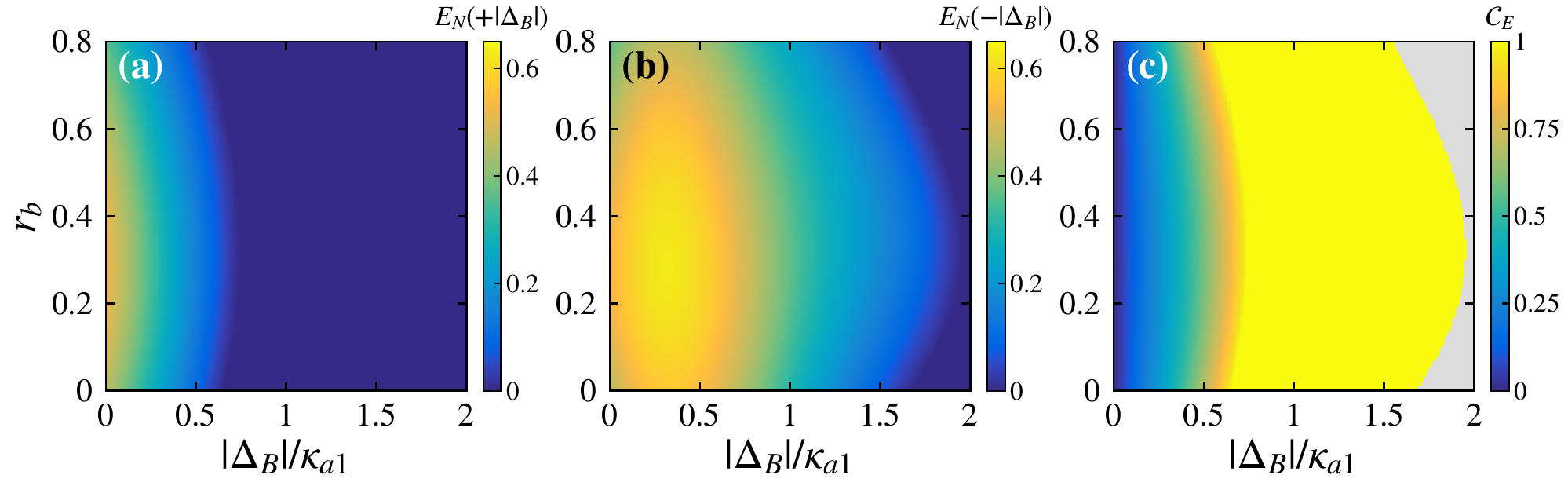}
	\caption{Steady-state magnon--magnon entanglement $E_N$ for
		(a) positive and (b) negative Barnett shifts, and
		(c) the bidirectional contrast ratio $\mathcal C_E$, versus
		$|\Delta_B|/\kappa_{a1}$ and $r_b$.
		Gray regions in (c) indicate vanishing entanglement for both
		rotation directions, where $\mathcal C_E$ is undefined.
		Other parameters are the same as in Fig.~\ref{fig2}.}
	\label{fig3}
\end{figure*}

To characterize the entanglement asymmetry under reversal of the Barnett
shift, we define the normalized contrast~\cite{Lu2025}
\begin{equation}
	\mathcal C_E=
	\frac{|E_N(\Delta_B>0)-E_N(\Delta_B<0)|}
	{E_N(\Delta_B>0)+E_N(\Delta_B<0)}.
	\label{eq:cfmm_entanglement_contrast}
\end{equation}
When the denominator is nonzero,
$\mathcal C_E=0$ indicates equal entanglement for the two signs, whereas
$\mathcal C_E=1$ means that entanglement exists for only one sign. The
contrast is undefined when both logarithmic negativities vanish.
Figures~\ref{fig3}(a) and~\ref{fig3}(b) show $E_N$ versus
$|\Delta_B|$ and $r_b$ for positive and negative Barnett shifts,
respectively.
For the parameters considered here, stronger entanglement is
obtained for $\Delta_B<0$ and persists over a wider range of
$|\Delta_B|$.
From Fig.~\ref{fig3}(c), we find that $\mathcal C_E$ reaches unity
where entanglement vanishes for $\Delta_B>0$ but remains finite
for $\Delta_B<0$.
This result shows that reversing the rotation direction can
switch magnon--magnon entanglement on and off.
The gray regions correspond to vanishing entanglement for both signs.
We also observe that increasing $r_b$ does not necessarily
enhance both entanglement and its normalized contrast.
Specifically, at $|\Delta_B|=0.3\kappa_{a1}$, increasing $r_b$
from 0 to 0.3 enhances $E_N$ for both signs, while
$\mathcal C_E$ decreases from approximately 0.40 to 0.35.

\begin{figure}[htbp]
	\centering
	\includegraphics[width=0.5\textwidth]
	{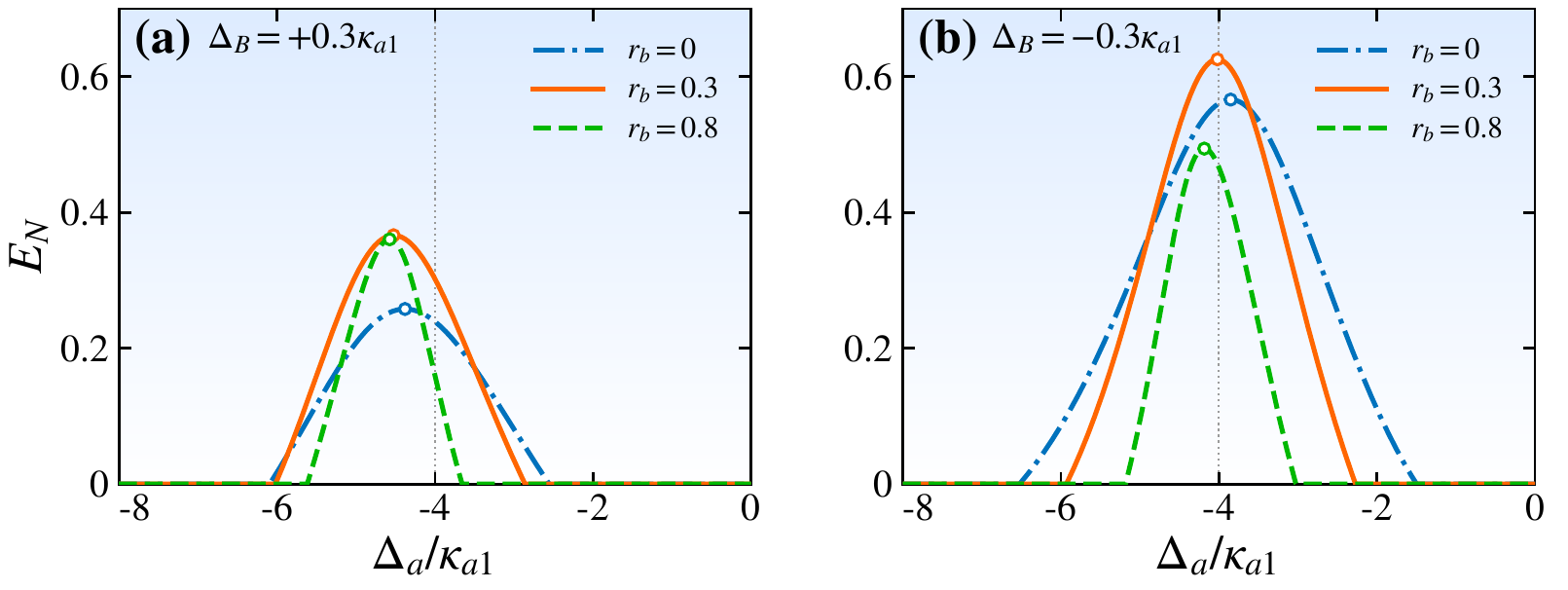}
	\caption{Steady-state magnon--magnon entanglement $E_N$ versus
		the common cavity detuning $\Delta_{a1}=\Delta_{a2}=\Delta_a$
		for (a) $\Delta_B=+0.3\kappa_{a1}$ and
		(b) $\Delta_B=-0.3\kappa_{a1}$.
		The curves correspond to $r_b=0,0.3,0.8$.
		Open circles mark the maxima, and vertical dotted lines indicate
		$\Delta_a=-4\kappa_{a1}$.
		Other parameters are the same as in Fig.~\ref{fig3}.}
	\label{fig4}
\end{figure}

Figure~\ref{fig4} shows $E_N$ versus the common cavity detuning
$\Delta_{a1}=\Delta_{a2}=\Delta_a$ for
$\Delta_B=\pm0.3\kappa_{a1}$.
The two signs of the Barnett shift yield different peak positions
and different intervals in which entanglement exists.
This behavior arises from the change in the magnon detuning
under reversal of $\Delta_B$.
In particular, for $r_b=0.8$ and
$\Delta_a=-5.4\kappa_{a1}$, entanglement exists only for
the positive Barnett shift.
At $\Delta_a=-3.4\kappa_{a1}$, the opposite behavior is observed:
entanglement survives only for the negative shift.
These results show that the rotation direction supporting
magnon--magnon entanglement can be selected by tuning
the cavity detuning.
Moreover, increasing $r_b$ from 0 to 0.3 raises the peak
entanglement in both cases, while narrowing the interval
of cavity detunings with nonzero entanglement.

\begin{figure}[htbp]
	\centering
	\includegraphics[width=0.5\textwidth]
	{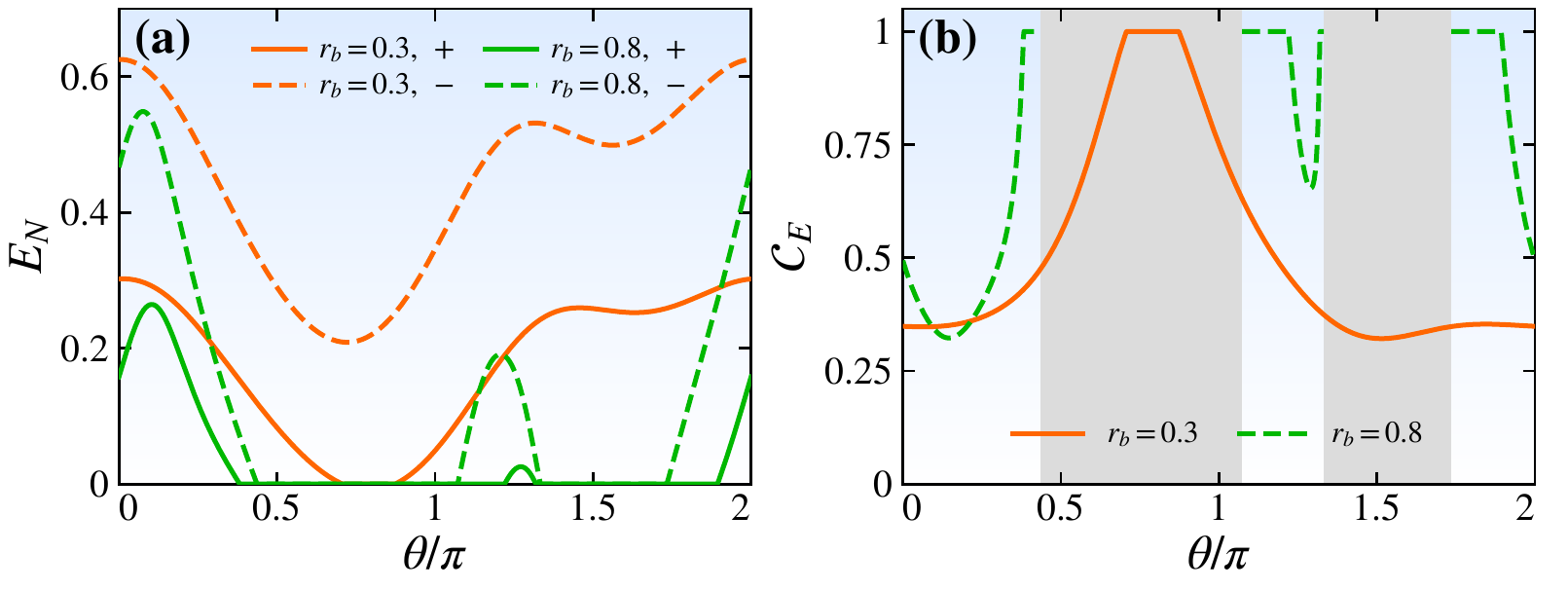}
	\caption{(a) Steady-state magnon--magnon entanglement $E_N$
		and (b) normalized contrast $\mathcal C_E$ versus the
		feedback phase $\theta$ for $r_b=0.3$ (orange) and
		$r_b=0.8$ (green), with $|\Delta_B|=0.3\kappa_{a1}$.
		Solid and dashed curves in panel (a) correspond to
		positive and negative Barnett shifts, respectively.
		Gray regions in panel (b) indicate undefined contrast
		at $r_b=0.8$, where entanglement vanishes for both signs.
		Other parameters are the same as in Fig.~\ref{fig4}.}
	\label{fig5}
\end{figure}

Figure~\ref{fig5}(a) shows that $E_N$ depends strongly
on $\theta$, particularly for $r_b=0.8$.
In this case, an appropriate feedback phase enhances
entanglement for both signs of $\Delta_B$ compared with
$\theta=0$, while the optimal phases are different.
Changing $\theta$ modifies the effective decay rate and
detuning of cavity $a_2$, as well as its squeezed-input
correlations, thereby affecting the transfer of quantum
correlations to the magnon modes.
From Fig.~\ref{fig5}(b), we find that $\mathcal C_E$
reaches unity within certain phase intervals.
For example, at $r_b=0.3$ and $\theta=0.8\pi$,
entanglement vanishes for $\Delta_B>0$ but remains finite
for $\Delta_B<0$.
Thus, the feedback phase can be tuned so that reversing
the rotation direction switches magnon--magnon
entanglement on and off.
The gray regions correspond to vanishing entanglement
for both signs at $r_b=0.8$, where $\mathcal C_E$
is undefined.

\begin{figure*}[htbp]
	\centering
	\includegraphics[width=1\textwidth]
	{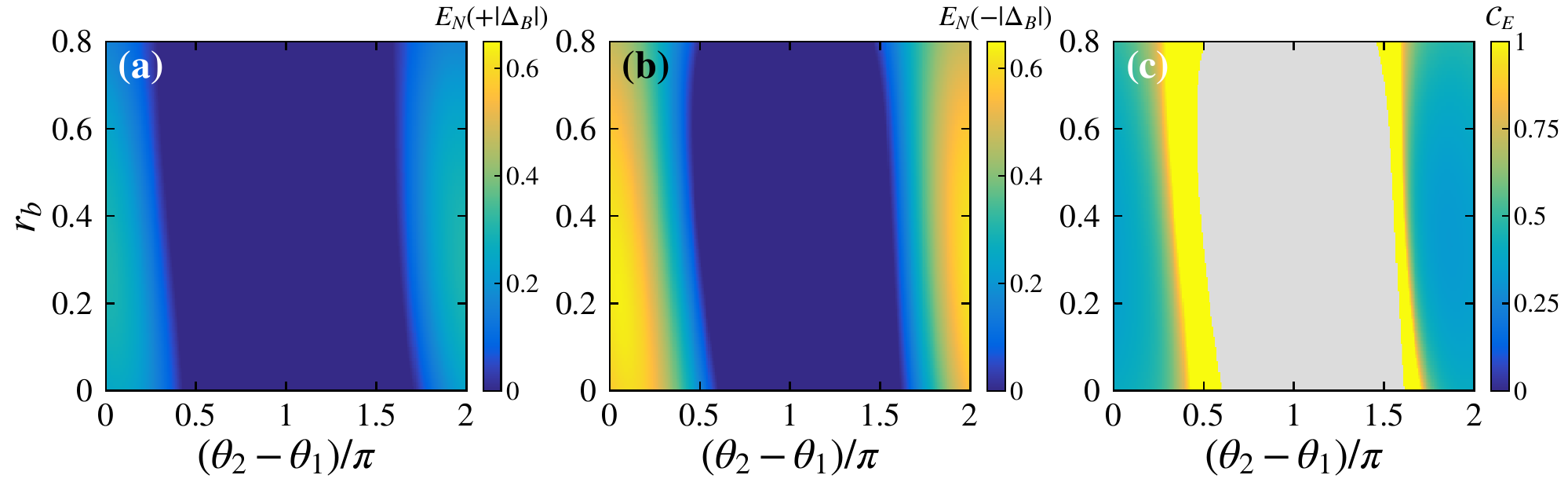}
	\caption{Steady-state magnon--magnon entanglement $E_N$
		for (a) $\Delta_B=+0.3\kappa_{a1}$ and
		(b) $\Delta_B=-0.3\kappa_{a1}$, and
		(c) normalized contrast $\mathcal C_E$, versus the
		relative squeezing phase $\theta_2-\theta_1$ and
		feedback reflectivity $r_b$.
		Here, $\theta_1=\theta=0$ and $R_1=R_2=0.9$.
		Gray regions in panel (c) indicate undefined contrast,
		where entanglement vanishes for both signs.
		Other parameters are the same as in Fig.~\ref{fig5}.}
	\label{fig6}
	
\end{figure*}

Next, we investigate the effect of the relative phase
between the two squeezed input fields, keeping the
feedback phase fixed at $\theta=0$.
Figures~\ref{fig6}(a) and~\ref{fig6}(b) show that
relatively strong entanglement is obtained when the
two squeezed fields are nearly in phase, whereas
entanglement vanishes around $\theta_2-\theta_1=\pi$.
With the squeezing factors fixed, the relative phase
affects the transfer of quantum correlations from the
two squeezed fields to the magnon modes.
For the parameters considered here, entanglement
persists over a wider phase interval for $\Delta_B<0$.
From Fig.~\ref{fig6}(c), we find regions where
$\mathcal C_E$ reaches unity.
For example, at $r_b=0.3$ and
$\theta_2-\theta_1=0.4\pi$, entanglement vanishes
for $\Delta_B>0$ but remains finite for $\Delta_B<0$,
with $E_N\simeq0.20$.
Compared with equal squeezing phases, the normalized
contrast increases from approximately 0.35 to 1,
while the entanglement for the negative shift decreases.
The gray regions correspond to vanishing entanglement
for both signs, where $\mathcal C_E$ is undefined.

\begin{figure}[htbp]
	\centering
	\includegraphics[width=0.5\textwidth]
	{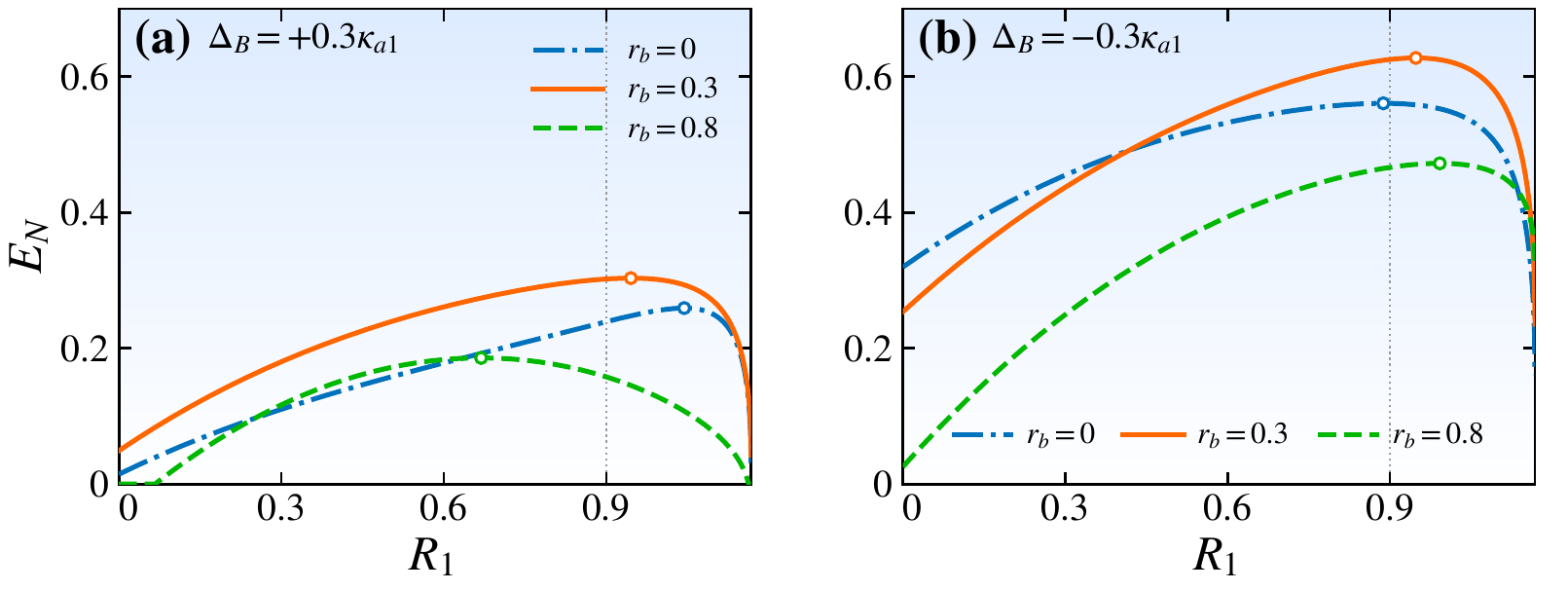}
	\caption{Steady-state magnon--magnon entanglement $E_N$
		versus the squeezing factor $R_1$ for
		(a) $\Delta_B=+0.3\kappa_{a1}$ and
		(b) $\Delta_B=-0.3\kappa_{a1}$.
		The second squeezing factor $R_2$ is determined by
		$\sinh^2R_1+\sinh^2R_2=2\sinh^2(0.9)$.
		The curves correspond to $r_b=0,0.3,0.8$.
		Open circles mark the maxima, and vertical dotted lines
		indicate equal squeezing, $R_1=R_2=0.9$.
		Other parameters are the same as in Fig.~\ref{fig6}.}
	\label{fig7}
\end{figure}

Then we investigate the effect of the two squeezing
factors while keeping $N_1+N_2$ fixed.
Figure~\ref{fig7} shows that $E_N$ first increases
and then decreases as $R_1$ increases, with $R_2$
decreasing under the constraint.
For all feedback reflectivities considered here,
the maximum entanglement is obtained when both
inputs are squeezed.
The optimal squeezing factors depend on both $r_b$
and the sign of the Barnett shift.
In particular, at $r_b=0.8$, the maxima for positive
and negative Barnett shifts occur at
$(R_1,R_2)\simeq(0.67,1.05)$ and $(0.99,0.79)$,
respectively.
For the positive shift, adjusting the squeezing factors
increases $E_N$ from approximately 0.158 at equal
squeezing to 0.186, without changing $N_1+N_2$.
The different optimal strengths reflect the unequal
contributions of the two squeezed inputs to the
magnon correlations.
Coherent feedback changes the effective decay rate
of cavity $a_2$, while the Barnett effect changes
the detuning of $m_1$, thereby affecting the transfer
of quantum correlations from the two input fields.
\begin{figure}[htbp]
	\centering
	\includegraphics[width=0.5\textwidth]
	{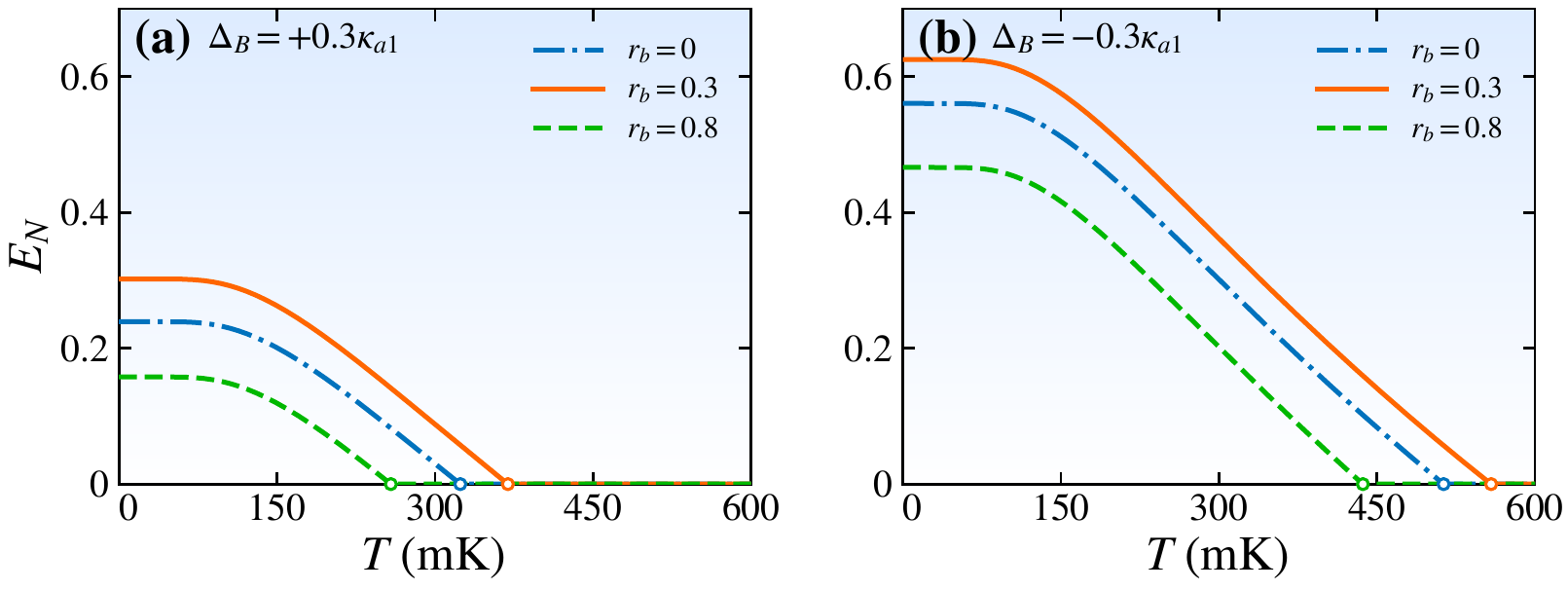}
	\caption{Steady-state magnon--magnon entanglement $E_N$
		versus the temperature $T$ of the two magnon baths for
		(a) $\Delta_B=+0.3\kappa_{a1}$ and
		(b) $\Delta_B=-0.3\kappa_{a1}$.
		The curves correspond to $r_b=0,0.3,0.8$.
		Open circles mark the critical temperatures at which
		entanglement vanishes.
		The squeezed inputs are kept fixed, and other parameters
		are the same as in Fig.~\ref{fig7}.}
	\label{fig8}
\end{figure}

Finally, we examine the robustness of nonreciprocal
magnon--magnon entanglement against thermal noise.
Figure~\ref{fig8} shows that $E_N$ decreases with
increasing temperature and vanishes above a critical
temperature.
This reduction arises from the increase in thermal
magnon occupations.
For the parameters considered here, entanglement
survives at higher temperatures for $\Delta_B<0$
than for $\Delta_B>0$.
Increasing $r_b$ from 0 to 0.3 raises the critical
temperature from approximately $324\,\mathrm{mK}$
to $369\,\mathrm{mK}$ for the positive Barnett shift,
and from $513\,\mathrm{mK}$ to $558\,\mathrm{mK}$
for the negative shift.
For $r_b=0.3$, temperatures between these two thresholds
yield entanglement only for $\Delta_B<0$, corresponding
to $\mathcal C_E=1$.
The critical temperatures decrease when $r_b$ is
further increased to 0.8, showing that stronger feedback
does not necessarily improve thermal robustness.

\section{Conclusion}\label{sec3}

We proposed a scheme for generating and controlling
remote nonreciprocal magnon--magnon entanglement in a
coupled cavity--magnon system with coherent feedback.
Quantum correlations from two independently squeezed
microwave fields are transferred to the magnon modes
through the coupled cavities.
The Barnett effect changes the detuning of the 
magnon mode, leading to different steady-state
entanglement for opposite rotation directions.
By appropriately tuning the cavity detuning and
feedback phase, entanglement can be obtained for one
rotation direction while it vanishes for the opposite
direction. Our results show that coherent feedback enhances the
entanglement at an appropriate reflectivity, whereas
further increasing the reflectivity can reduce it.
The relative squeezing phase and the two squeezing
factors provide additional control of the entanglement.
In particular, an appropriate choice of unequal
squeezing factors improves the entanglement.
With the squeezed inputs kept fixed, moderate feedback
also increases the critical temperature above which
magnon--magnon entanglement vanishes.
These findings provide a controllable scheme for
preparing remote nonreciprocal entanglement in
cavity magnonic systems.

\section{Acknowledgements}
Mei Zhang thanks Junzhong Yang for helpful discussions. Jinhao Jia thanks
Lijun He, Fan Zhou and Hejin Lv for helpful discussions. Yingru Li thanks Xinchun Li and Juan Yao for helpful discussions. This work was supported by the National Natural Science Foundation of China under Grant No.
11475021 and the National Key Basic Research Program of
China under Grant No. 2013CB922000.

\bibliographystyle{unsrtnat}

\bibliography{refs}

@article{Horodecki2009,
	title = {Quantum entanglement},
	author = {R. Horodecki and P. Horodecki and M. Horodecki and K. Horodecki},
	journal = {Rev. Mod. Phys.},
	volume = {81},
	issue = {2},
	pages = {865--942},
	numpages = {0},
	year = {2009},
	month = {Jun},
	publisher = {American Physical Society},
	doi = {10.1103/RevModPhys.81.865},
}

@article{Tabuchi2014,
	title = {Hybridizing Ferromagnetic Magnons and Microwave Photons in the Quantum Limit},
	author = {Y. Tabuchi and S. Ishino and T. Ishikawa and R. Yamazaki and K. Usami and Y. Nakamura},
	journal = {Phys. Rev. Lett.},
	volume = {113},
	issue = {8},
	pages = {083603},
	numpages = {5},
	year = {2014},
	month = {Aug},
	publisher = {American Physical Society},
	doi = {10.1103/PhysRevLett.113.083603},
}

@article{Huebl2013,
	title = {High Cooperativity in Coupled Microwave Resonator Ferrimagnetic Insulator Hybrids},
	author = {H. Huebl and C. W. Zollitsch and J. Lotze and F. Hocke and M. Greifenstein and A. Marx and R. Gross and S. T. B. Goennenwein},
	journal = {Phys. Rev. Lett.},
	volume = {111},
	issue = {12},
	pages = {127003},
	numpages = {5},
	year = {2013},
	month = {Sep},
	publisher = {American Physical Society},
	doi = {10.1103/PhysRevLett.111.127003},
}

@article{Zhang2019,
	title = {Quantum entanglement between two magnon modes via Kerr nonlinearity driven far from equilibrium},
	author = {Z. Zhang and M. O. Scully and G. S. Agarwal},
	journal = {Phys. Rev. Research},
	volume = {1},
	issue = {2},
	pages = {023021},
	numpages = {7},
	year = {2019},
	month = {Sep},
	publisher = {American Physical Society},
	doi = {10.1103/PhysRevResearch.1.023021},
}

@article{Li2018,
	title = {Magnon-Photon-Phonon Entanglement in Cavity Magnomechanics},
	author = {J. Li and S.-Y. Zhu and G. S. Agarwal},
	journal = {Phys. Rev. Lett.},
	volume = {121},
	issue = {20},
	pages = {203601},
	numpages = {6},
	year = {2018},
	month = {Nov},
	publisher = {American Physical Society},
	doi = {10.1103/PhysRevLett.121.203601},
}

@article{Yu2020,
	doi = {10.1088/1361-6455/ab68b5},
	year = {2020},
	month = {feb},
	publisher = {IOP Publishing},
	volume = {53},
	number = {6},
	pages = {065402},
	author = {M. Yu and S.-Y. Zhu and J. Li},
	title = {Macroscopic entanglement of two magnon modes via quantum correlated microwave fields},
	journal = {J. Phys. B: At. Mol. Opt. Phys.},
}

@article{Yuan2025,
	title = {Manipulation of asymmetric steering and periodic entanglement in cavity magnetomechanics induced by chiral exceptional points},
	author = {Z.-H. Yuan and Y. Xia and Y.-Y. Jiang and J. Song},
	journal = {Phys. Rev. B},
	volume = {111},
	issue = {1},
	pages = {014433},
	numpages = {12},
	year = {2025},
	month = {Jan},
	publisher = {American Physical Society},
	doi = {10.1103/PhysRevB.111.014433},
}

@article{Guo2025,
	title = {Nonreciprocal macroscopic entanglement between two magnon modes induced by Kerr nonlinearity},
	author = {Q. Guo and J.-X. Wang and C.-H. Bai and Y. Zhang and G. Li and T. Zhang},
	journal = {Phys. Rev. B},
	volume = {112},
	issue = {14},
	pages = {144422},
	numpages = {9},
	year = {2025},
	month = {Oct},
	publisher = {American Physical Society},
	doi = {10.1103/343d-t8pv},
}

@article{Zhang2022,
	author = {W. Zhang and T. Wang and X. Han and S. Zhang and H.-F. Wang},
	journal = {Opt. Express},
	number = {7},
	pages = {10969--10980},
	publisher = {Optica Publishing Group},
	title = {Quantum entanglement and one-way steering in a cavity magnomechanical system via a squeezed vacuum field},
	volume = {30},
	month = {Mar},
	year = {2022},
	doi = {10.1364/OE.453787},
}

@article{Lu2024,
	author = {T.-X. Lu and Z.-S. Li and B. Yin and J. Wang and X. Xiao and H. Jing},
	journal = {Opt. Express},
	number = {27},
	pages = {48302--48314},
	publisher = {Optica Publishing Group},
	title = {Magnetic-field-direction-controlled slow light and second-order sidebands in a cavity-magnon optomechanical system},
	volume = {32},
	month = {Dec},
	year = {2024},
	doi = {10.1364/OE.546225},
}

@article{peng2025,
	title = {Symmetric vs asymmetric magnon pairing: Comparative study of quantum entanglement and synchronization in the cavity-magnon system},
	author = {J.-X. Peng and S. K. Singh and N. Akhtar and Z. Gu and C. Wu and J.-F. Li},
	journal = {Phys. Rev. A},
	volume = {112},
	issue = {4},
	pages = {042430},
	numpages = {10},
	year = {2025},
	month = {Oct},
	publisher = {American Physical Society},
	doi = {10.1103/7hhn-zypy},
}

@article{Kani2022,
	title = {Magnonic Einstein--de Haas Effect: Ultrafast Rotation of Magnonic Microspheres},
	author = {A. Kani and F. Quijandr{\'{\i}}a and J. Twamley},
	journal = {Phys. Rev. Lett.},
	volume = {129},
	issue = {25},
	pages = {257201},
	numpages = {7},
	year = {2022},
	month = {Dec},
	publisher = {American Physical Society},
	doi = {10.1103/PhysRevLett.129.257201},
}

@article{Ono2015,
	title = {Barnett effect in paramagnetic states},
	author = {M. Ono and H. Chudo and K. Harii and S. Okayasu and M. Matsuo and J. Ieda and R. Takahashi and S. Maekawa and E. Saitoh},
	journal = {Phys. Rev. B},
	volume = {92},
	issue = {17},
	pages = {174424},
	numpages = {4},
	year = {2015},
	month = {Nov},
	publisher = {American Physical Society},
	doi = {10.1103/PhysRevB.92.174424},
}

@article{Bretzel2009,
	author  = {S. Bretzel and G. E. W. Bauer and Y. Tserkovnyak and A. Brataas},
	title   = {{Barnett} effect in thin magnetic films and nanostructures},
	journal = {Appl. Phys. Lett.},
	volume  = {95},
	number  = {12},
	pages   = {122504},
	year    = {2009},
	doi     = {10.1063/1.3232221}
}

@article{Gong2026Remote,
  author  = {Y. Gong and Y.-X. Cheng and W. Xiong and J. Chen},
  title   = {Remote magnon-magnon entanglement enhanced by squeezed-field interference},
  journal = {APS Open Sci.},
  volume  = {1},
  pages   = {000027},
  year    = {2026},
  doi     = {10.1103/m3z7-z9t4}
}

@article{Jia2026Coherent,
  author  = {J. Jia and Y. Li and R. Liang and M. Zhang},
  title   = {Coherent control of optomechanical entanglement and steering via dual parametric amplification},
  journal = {Phys. Rev. A},
  volume  = {114},
  pages   = {033514},
  year    = {2026},
  doi     = {10.1103/nrxl-g4lt}
}

@article{Jia2026,
  author  = {J. Jia and Y. Li and J. Huang and M. Zhang},
  title   = {Nonreciprocal quantum coherence in a cavity magnomechanical system via the {Barnett} effect},
  journal = {Phys. Rev. A},
  volume  = {113},
  pages   = {013719},
  year    = {2026},
  doi     = {10.1103/vmwj-9qm8}
}

@article{Lu2025,
	title = {Nonreciprocal entanglement in cavity magnomechanics via the Barnett effect},
	author = {T.-X. Lu and Z.-S. Li and L.-S. Chen and Y. Wang and X. Xiao and H. Jing},
	journal = {Phys. Rev. A},
	volume = {111},
	issue = {1},
	pages = {013713},
	numpages = {10},
	year = {2025},
	month = {Jan},
	publisher = {American Physical Society},
	doi = {10.1103/PhysRevA.111.013713},
}

@article{Hu2026,
	author  = {W. Y. Hu and H. Yi and H. N. Liu and S. Y. Ge and C. Shang and Y.-H. Zhou and H. Z. Shen},
	title   = {Nonreciprocal coupling induced non-{Hermitian} entanglement in optomechanical systems},
	journal = {Laser Photonics Rev.},
	pages   = {e71950},
	year    = {2026},
	doi     = {10.1002/lpor.71950}
}

@article{Imara2016,
	title = {Nonreciprocal macroscopic entanglement through magnon squeezing in cavity magnomechanics},
	author = {Z. Imara and K. {El Anouz} and {\.{I}}. Demir and A. {El Allati}},
	journal = {Phys. Rev. A},
	volume = {113},
	issue = {5},
	pages = {053717},
	numpages = {10},
	year = {2026},
	month = {May},
	publisher = {American Physical Society},
	doi = {10.1103/rl1p-vgj1},
}

@article{Yang2026,
	title = {Simultaneous nonreciprocal unconventional photon blockades via two degenerate optical parametric amplifiers in spinning resonators},
	author = {J. X. Yang and Z.-L. Zhao and Z. Wang and T. Z. Luan and S. Xu and C. Shang and Y.-H. Zhou and H. Z. Shen},
	journal = {Phys. Rev. A},
	volume = {114},
	issue = {2},
	pages = {023703},
	numpages = {23},
	year = {2026},
	month = {Aug},
	publisher = {American Physical Society},
	doi = {10.1103/3brk-k1l9},
}

@article{Chen2024,
	title = {Nonreciprocal photon-phonon entanglement in Kerr-modified spinning cavity magnomechanics},
	author = {J. Chen and X.-G. Fan and W. Xiong and D. Wang and L. Ye},
	journal = {Phys. Rev. A},
	volume = {109},
	issue = {4},
	pages = {043512},
	numpages = {9},
	year = {2024},
	month = {Apr},
	publisher = {American Physical Society},
	doi = {10.1103/PhysRevA.109.043512},
}

@article{Chen2023,
	title = {Nonreciprocal entanglement in cavity-magnon optomechanics},
	author = {J. Chen and X.-G. Fan and W. Xiong and D. Wang and L. Ye},
	journal = {Phys. Rev. B},
	volume = {108},
	issue = {2},
	pages = {024105},
	numpages = {8},
	year = {2023},
	month = {Jul},
	publisher = {American Physical Society},
	doi = {10.1103/PhysRevB.108.024105},
}

@article{Lloyd2000,
	title = {Coherent quantum feedback},
	author = {S. Lloyd},
	journal = {Phys. Rev. A},
	volume = {62},
	issue = {2},
	pages = {022108},
	numpages = {12},
	year = {2000},
	month = {Jul},
	publisher = {American Physical Society},
	doi = {10.1103/PhysRevA.62.022108},
}

@article{Schmid2022,
	title = {Coherent Feedback Cooling of a Nanomechanical Membrane with Atomic Spins},
	author = {G.-L. Schmid and C. T. Ngai and M. Ernzer and M. {Bosch Aguilera} and T. M. Karg and P. Treutlein},
	journal = {Phys. Rev. X},
	volume = {12},
	issue = {1},
	pages = {011020},
	numpages = {13},
	year = {2022},
	month = {Jan},
	publisher = {American Physical Society},
	doi = {10.1103/PhysRevX.12.011020},
}

@article{VidalWerner2002,
  author  = {G. Vidal and R. F. Werner},
  title   = {Computable measure of entanglement},
  journal = {Phys. Rev. A},
  volume  = {65},
  pages   = {032314},
  year    = {2002},
  doi     = {10.1103/PhysRevA.65.032314}
}

@article{Plenio2005,
  author  = {M. B. Plenio},
  title   = {Logarithmic negativity: A full entanglement monotone that is not convex},
  journal = {Phys. Rev. Lett.},
  volume  = {95},
  pages   = {090503},
  year    = {2005},
  doi     = {10.1103/PhysRevLett.95.090503}
}

@article{Simon2000,
  author  = {R. Simon},
  title   = {{Peres-Horodecki} separability criterion for continuous variable systems},
  journal = {Phys. Rev. Lett.},
  volume  = {84},
  pages   = {2726},
  year    = {2000},
  doi     = {10.1103/PhysRevLett.84.2726}
}

@article{Hisatomi2016,
	title = {Bidirectional conversion between microwave and light via ferromagnetic magnons},
	author = {R. Hisatomi and A. Osada and Y. Tabuchi and T. Ishikawa and A. Noguchi and R. Yamazaki and K. Usami and Y. Nakamura},
	journal = {Phys. Rev. B},
	volume = {93},
	issue = {17},
	pages = {174427},
	numpages = {13},
	year = {2016},
	month = {May},
	publisher = {American Physical Society},
	doi = {10.1103/PhysRevB.93.174427},
}

@article{
	Dany2017,
	author = {D. Lachance-Quirion and Y. Tabuchi and S. Ishino and A. Noguchi and T. Ishikawa and R. Yamazaki and Y. Nakamura},
	title = {Resolving quanta of collective spin excitations in a millimeter-sized ferromagnet},
	journal = {Sci. Adv.},
	volume = {3},
	number = {7},
	pages = {e1603150},
	year = {2017},
	doi = {10.1126/sciadv.1603150},
}

@article{Wolski2020,
	title = {Dissipation-Based Quantum Sensing of Magnons with a Superconducting Qubit},
	author = {S. P. Wolski and D. Lachance-Quirion and Y. Tabuchi and S. Kono and A. Noguchi and K. Usami and Y. Nakamura},
	journal = {Phys. Rev. Lett.},
	volume = {125},
	issue = {11},
	pages = {117701},
	numpages = {6},
	year = {2020},
	month = {Sep},
	publisher = {American Physical Society},
	doi = {10.1103/PhysRevLett.125.117701},
}

@article{Babak2022,
	title = {Cavity magnonics},
	journal = {Phys. Rep.},
	volume = {979},
	pages = {1-61},
	year = {2022},
	note = {Cavity Magnonics},
	issn = {0370-1573},
	doi = {10.1016/j.physrep.2022.06.001},
	author = {B. {Zare Rameshti} and S. {Viola Kusminskiy} and J. A. Haigh and K. Usami and D. Lachance-Quirion and Y. Nakamura and C.-M. Hu and H. X. Tang and G. E. W. Bauer and Y. M. Blanter}}

@article{Laure2021,
	author = {L. {Mercier de L{\'e}pinay} and C. F. Ockeloen-Korppi and M. J. Woolley and M. A. Sillanp{\"a}{\"a}},
	title = {Quantum mechanics--free subsystem with mechanical oscillators},
	journal = {Science},
	volume = {372},
	pages = {625--629},
	year = {2021},
	doi = {10.1126/science.abf5389}
}

@article{Shlomi2021,
	author = {S. Kotler and G. A. Peterson and E. Shojaee and F. Lecocq and K. Cicak and A. Kwiatkowski and S. Geller and S. Glancy and E. Knill and R. W. Simmonds and J. Aumentado and J. D. Teufel},
	title = {Direct observation of deterministic macroscopic entanglement},
	journal = {Science},
	volume = {372},
	pages = {622--625},
	year = {2021},
	doi = {10.1126/science.abf2998}
}

@article{Tao2020,
	author = {T. Li and Z. Wang and K. Xia},
	title = {Multipartite quantum entanglement creation for distant stationary systems},
	journal = {Opt. Express},
	volume = {28},
	pages = {1316--1329},
	year = {2020},
	doi = {10.1364/OE.383152}
}

@article{lv2018,
	author = {X.-Y. L\"u and G.-L. Zhu and L.-L. Zheng and Y. Wu},
	title = {Entanglement and quantum superposition induced by a single photon},
	journal = {Phys. Rev. A},
	volume = {97},
	pages = {033807},
	year = {2018},
	doi = {10.1103/PhysRevA.97.033807}
}

@article{Pezz2018,
	author = {L. Pezz\`e and A. Smerzi and M. K. Oberthaler and R. Schmied and P. Treutlein},
	title = {Quantum metrology with nonclassical states of atomic ensembles},
	journal = {Rev. Mod. Phys.},
	volume = {90},
	pages = {035005},
	year = {2018},
	doi = {10.1103/RevModPhys.90.035005}
}

@article{zhang2015,
	author = {Z. Zhang and S. Mouradian and F. N. C. Wong and J. H. Shapiro},
	title = {Entanglement-Enhanced Sensing in a Lossy and Noisy Environment},
	journal = {Phys. Rev. Lett.},
	volume = {114},
	pages = {110506},
	year = {2015},
	doi = {10.1103/PhysRevLett.114.110506}
}

@article{Graham2022,
	author = {T. M. Graham and Y. Song and J. Scott and C. Poole and L. Phuttitarn and K. Jooya and P. Eichler and X. Jiang and A. Marra and B. Grinkemeyer and M. Kwon and M. Ebert and J. Cherek and M. T. Lichtman and M. Gillette and J. Gilbert and D. Bowman and T. Ballance and C. Campbell and E. D. Dahl and O. Crawford and N. S. Blunt and B. Rogers and T. Noel and M. Saffman},
	title = {Multi-Qubit Entanglement and Algorithms on a Neutral-Atom
	Quantum Computer},
	journal = {Nature},
	volume = {604},
	pages = {457--462},
	year = {2022},
	doi = {10.1038/s41586-022-04603-6}
}

@article{Bartolucci2023,
	author = {S. Bartolucci and P. Birchall and H. Bomb{\'\i}n and H. Cable and C. Dawson and M. Gimeno-Segovia and E. Johnston and K. Kieling and N. Nickerson and M. Pant and F. Pastawski and T. Rudolph and C. Sparrow},
	title = {Fusion-Based Quantum Computation},
	journal = {Nat. Commun.},
	volume = {14},
	pages = {912},
	year = {2023},
	doi = {10.1038/s41467-023-36493-1}
}

@article{Ursin2007,
	author = {R. Ursin and F. Tiefenbacher and T. Schmitt-Manderbach and H. Weier and T. Scheidl and M. Lindenthal and B. Blauensteiner and T. Jennewein and J. Perdigues and P. Trojek and B. {\"O}mer and M. F{\"u}rst and M. Meyenburg and J. Rarity and Z. Sodnik and C. Barbieri and H. Weinfurter and A. Zeilinger},
	title = {Entanglement-Based Quantum Communication over 144 Km},
	journal = {Nat. Phys.},
	volume = {3},
	pages = {481--486},
	year = {2007},
	doi = {10.1038/nphys629}
}

@article{Piveteau2022,
	author = {A. Piveteau and J. Pauwels and E. H{\aa}kansson and S. Muhammad and M. Bourennane and A. Tavakoli},
	title = {Entanglement-Assisted Quantum Communication with Simple
	Measurements},
	journal = {Nat. Commun.},
	volume = {13},
	pages = {7878},
	year = {2022},
	doi = {10.1038/s41467-022-33922-5}
}

@article{Ji2023,
	title = {Kerr nonlinearity induced strong spin-magnon coupling},
	author = {F.-Z. Ji and J.-H. An},
	journal = {Phys. Rev. B},
	volume = {108},
	issue = {18},
	pages = {L180409},
	numpages = {7},
	year = {2023},
	month = {Nov},
	publisher = {American Physical Society},
	doi = {10.1103/PhysRevB.108.L180409},
}

@article{Li2019,
	title = {Squeezed states of magnons and phonons in cavity magnomechanics},
	author = {J. Li and S.-Y. Zhu and G. S. Agarwal},
	journal = {Phys. Rev. A},
	volume = {99},
	issue = {2},
	pages = {021801(R)},
	numpages = {6},
	year = {2019},
	month = {Feb},
	publisher = {American Physical Society},
	doi = {10.1103/PhysRevA.99.021801},
}

@article{Xiong2022,
	title = {Strong long-range spin-spin coupling via a Kerr magnon interface},
	author = {W. Xiong and M. Tian and G.-Q. Zhang and J. Q. You},
	journal = {Phys. Rev. B},
	volume = {105},
	issue = {24},
	pages = {245310},
	numpages = {7},
	year = {2022},
	month = {Jun},
	publisher = {American Physical Society},
	doi = {10.1103/PhysRevB.105.245310},
}

@article{Zhang2021,
	title = {Parity-symmetry-breaking quantum phase transition via parametric drive in a cavity magnonic system},
	author = {G.-Q. Zhang and Z. Chen and W. Xiong and C.-H. Lam and J. Q. You},
	journal = {Phys. Rev. B},
	volume = {104},
	issue = {6},
	pages = {064423},
	numpages = {10},
	year = {2021},
	month = {Aug},
	publisher = {American Physical Society},
	doi = {10.1103/PhysRevB.104.064423},
}

@article{Zhang2016,
	title   = {Cavity magnomechanics},
	author  = {X. Zhang and C.-L. Zou and L. Jiang and H. X. Tang},
	journal = {Sci. Adv.},
	volume  = {2},
	number  = {3},
	pages   = {e1501286},
	year    = {2016},
	doi     = {10.1126/sciadv.1501286}
}

@article{Wiseman1994,
	author  = {H. M. Wiseman and G. J. Milburn},
	title   = {All-optical versus electro-optical quantum-limited feedback},
	journal = {Phys. Rev. A},
	volume  = {49},
	number  = {5},
	pages   = {4110--4125},
	year    = {1994},
	doi     = {10.1103/PhysRevA.49.4110}
}

@article{Gough2009,
	author  = {J. E. Gough and S. Wildfeuer},
	title   = {Enhancement of field squeezing using coherent feedback},
	journal = {Phys. Rev. A},
	volume  = {80},
	number  = {4},
	pages   = {042107},
	year    = {2009},
	doi     = {10.1103/PhysRevA.80.042107}
}

@article{Sudhir2017,
	author  = {V. Sudhir and D. J. Wilson and R. Schilling and H. Sch{\"u}tz and S. A. Fedorov and A. H. Ghadimi and A. Nunnenkamp and T. J. Kippenberg},
	title   = {Appearance and disappearance of quantum correlations in
	measurement-based feedback control of a mechanical oscillator},
	journal = {Phys. Rev. X},
	volume  = {7},
	number  = {1},
	pages   = {011001},
	year    = {2017},
	doi     = {10.1103/PhysRevX.7.011001}
}

@article{Kerckhoff2013,
	author  = {J. Kerckhoff and R. W. Andrews and H. S. Ku and W. F. Kindel and K. Cicak and R. W. Simmonds and K. W. Lehnert},
	title   = {Tunable coupling to a mechanical oscillator circuit
	using a coherent feedback network},
	journal = {Phys. Rev. X},
	volume  = {3},
	number  = {2},
	pages   = {021013},
	year    = {2013},
	doi     = {10.1103/PhysRevX.3.021013}
}

@article{Harwood2021,
	author  = {A. Harwood and M. Brunelli and A. Serafini},
	title   = {Cavity optomechanics assisted by optical coherent feedback},
	journal = {Phys. Rev. A},
	volume  = {103},
	number  = {2},
	pages   = {023509},
	year    = {2021},
	doi     = {10.1103/PhysRevA.103.023509}
}

@article{Kerckhoff2010,
	author  = {J. Kerckhoff and H. I. Nurdin and D. S. Pavlichin and H. Mabuchi},
	title   = {Designing quantum memories with embedded control:
	photonic circuits for autonomous quantum error correction},
	journal = {Phys. Rev. Lett.},
	volume  = {105},
	number  = {4},
	pages   = {040502},
	year    = {2010},
	doi     = {10.1103/PhysRevLett.105.040502}
}

@article{Kerckhoff2011,
	author  = {J. Kerckhoff and D. S. Pavlichin and H. Chalabi and H. Mabuchi},
	title   = {Design of nanophotonic circuits for autonomous
	subsystem quantum error correction},
	journal = {New J. Phys.},
	volume  = {13},
	number  = {5},
	pages   = {055022},
	year    = {2011},
	doi     = {10.1088/1367-2630/13/5/055022}
}

@article{Cramer2016,
	author  = {J. Cramer and N. Kalb and M. A. Rol and B. Hensen and M. S. Blok and M. Markham and D. J. Twitchen and R. Hanson and T. H. Taminiau},
	title   = {Repeated quantum error correction on a continuously
	encoded qubit by real-time feedback},
	journal = {Nat. Commun.},
	volume  = {7},
	pages   = {11526},
	year    = {2016},
	doi     = {10.1038/ncomms11526}
}

@article{Asjad2016,
	author  = {M. Asjad and P. Tombesi and D. Vitali},
	title   = {Feedback control of two-mode output entanglement
	and steering in cavity optomechanics},
	journal = {Phys. Rev. A},
	volume  = {94},
	number  = {5},
	pages   = {052312},
	year    = {2016},
	doi     = {10.1103/PhysRevA.94.052312}
}

@article{Li2017,
	author  = {J. Li and G. Li and S. Zippilli and D. Vitali and T. Zhang},
	title   = {Enhanced entanglement of two different mechanical
	resonators via coherent feedback},
	journal = {Phys. Rev. A},
	volume  = {95},
	number  = {4},
	pages   = {043819},
	year    = {2017},
	doi     = {10.1103/PhysRevA.95.043819}
}

@article{Peng2023,
	author  = {R. Peng and C. Zhao and Z. Yang and J. Yang and L. Zhou},
	title   = {Enhancement of mechanical entanglement and asymmetric
	steering with coherent feedback},
	journal = {Phys. Rev. A},
	volume  = {107},
	number  = {1},
	pages   = {013507},
	year    = {2023},
	doi     = {10.1103/PhysRevA.107.013507}
}

@article{Zhou2024,
	author  = {B.-Y. Zhou and N. Pan and L. Yu and G.-X. Li},
	title   = {Markovian-feedback-controlled entanglement and
	{Einstein-Podolsky-Rosen} steering in a nondegenerate
	optical parametric oscillator},
	journal = {Phys. Rev. A},
	volume  = {110},
	number  = {3},
	pages   = {032401},
	year    = {2024},
	doi     = {10.1103/PhysRevA.110.032401}
}

@article{Jiao2026,
	author  = {Y.-F. Jiao and R.-C. Wang and J.-X. Liu and H. Zhang and Y.-C. Liang and Y. Wang and L.-M. Kuang and H. Jing},
	title   = {Enhancing optomechanical entanglement and mechanical squeezing
	by the synergistic effect of quadratic optomechanical coupling
	and coherent feedback},
	journal = {Adv. Quantum Technol.},
	volume  = {9},
	number  = {7},
	pages   = {e70354},
	year    = {2026},
	doi     = {10.1002/qute.70354}
}

@article{Amazioug2025,
	author  = {M. Amazioug and J.-X. Peng and D. Dutykh and M. Asjad},
	title   = {Emergence and enhancement of feedback control-induced
	quantum entanglement},
	journal = {Eur. Phys. J. Plus},
	volume  = {140},
	pages   = {132},
	year    = {2025},
	doi     = {10.1140/epjp/s13360-024-05937-y}
}

@article{Liu2025,
	author  = {G.-Q. Liu and C.-Z. Gao and N. Wang and L. Yu and A.-D. Zhu},
	title   = {Entanglement enhancement by the synergistic effect of
	optical parametric amplification and coherent feedback
	in cavity magnonics},
	journal = {Adv. Quantum Technol.},
	volume  = {8},
	number  = {10},
	pages   = {e2500088},
	year    = {2025},
	doi     = {10.1002/qute.202500088}
}

@article{Gao2025,
	author  = {C.-Z. Gao and G.-Q. Liu and N. Wang and L. Yu and A.-D. Zhu},
	title   = {Nonreciprocal entanglement in spinning cavity
	magnomechanical system with coherent feedback loop},
	journal = {Adv. Quantum Technol.},
	volume  = {8},
	number  = {5},
	pages   = {2500063},
	year    = {2025},
	doi     = {10.1002/qute.202500063}
}

@article{Lin2025,
	author  = {Y.-H. Lin and R.-C. Yang and Z.-B. Yang and H.-Y. Liu},
	title   = {Enhancement of microwave entanglement via coherent
	quantum feedback with cavity magnonics},
	journal = {Adv. Quantum Technol.},
	volume  = {8},
	number  = {11},
	pages   = {e2500296},
	year    = {2025},
	doi     = {10.1002/qute.202500296}
}

\end{document}